# Position Measurement of Multiple Microparticles in Hollow-Core Photonic Crystal Fiber by Coherent Optical Frequency Domain Reflectometry

**Jasper Podschus[1], Max Koeppel[1], Bernhard Schmauss[1,2]**
[1]*Institute of Microwaves and Photonics, Friedrich-Alexander-Universität Erlangen-Nürnberg (FAU),
Cauerstr. 9, 91058 Erlangen, Germany*
*jasper.podschus@fau.de*

**Abhinav Sharma[2], Sanju Sundaramahalingam[2], Shangran Xie[2], Philip St.J. Russell[2,3]**
[2]*Max Planck Institute for the Science of Light and [3]Department of Physics, FAU,
Staudtstr. 2, 91058 Erlangen, Germany*
*shangran.xie@mpl.mpg.de*

**Abstract:** Flying particle sensors in hollow-core photonic crystal fibers require accurate localization of the optically trapped microparticles. We report position measurement to micrometer-resolution, using optical frequency domain reflectometry, of two 1.65-µm-diameter polystyrene particles. © 2020 The Author(s)

## 1. Introduction

Microparticles optically trapped within the hollow-core (HC) of a photonic crystal fiber (PCF) form the basis of a new type of distributed fiber sensor [1]. These "flying particle sensors" allow multiple physical quantities to be monitored by tracking the response of trapped particles to a measurand, and offer a resolution limited only by the particle size and the precision with which its position can be measured. To measure a physical quantity at different positions, the trapped particle is propelled along the HC-PCF by optical scattering forces. Multi-parameter sensing requires the simultaneous trapping and localization of multiple particles with different physical properties. Although laser Doppler velocimetry potentially offers a position resolution of half the vacuum wavelength [1], it is difficult to map the absolute position of particles. A camera-based approach is limited to small lengths of fiber lying within the field of view of the camera.

Optical frequency domain reflectometry (OFDR) provides a means of overcoming these limitations, while permitting the position of multiple particles to be measured to a precision inversely proportional to the measurement bandwidth of the system. We previously reported particle position measurement with a standard deviation of ~140 µm using incoherent OFDR with a bandwidth of ~5 GHz [2]. Coherent OFDR (COFDR) with a bandwidth of ~6 THz was subsequently used to locate a 15 µm polystyrene particle in a HC-PCF with a standard deviation of ~10 µm [3].

Here we report for the first time the use of COFDR to efficiently measure the position of two polystyrene particles (diameters as small as 1.65 µm) with a spatial resolution in the µm range. The results are of high relevance in multi-parameter sensing using chains of optically trapped particles each with different sensing properties.

## 2. Measurement setup

The experimental setup is illustrated in Fig. 1. The laser beams from a COFDR system (1525-1565 nm) and the trapping laser (1064 nm) are combined at a dichroic mirror and delivered to the HC-PCF.

### 2.1. COFDR system

The COFDR system [3] was based on a single-mode fiber-coupled Mach-Zehnder interferometer and a tunable laser source (Luna Phoenix 1400), as shown within the blue-dashed box in Fig. 1. The laser output (10 mW power) was swept from 1525 nm to 1565 nm, corresponding to approximately 5 THz. This frequency-tuned laser light was split into a probe and reference path, the probe light being coupled into free space using a collimator, combined with a trapping beam using a dichroic mirror and coupled into the HC-PCF. The backscattered probe signal containing the reflection from a trapped particle interferes with the reference signal and is detected by a photodiode.

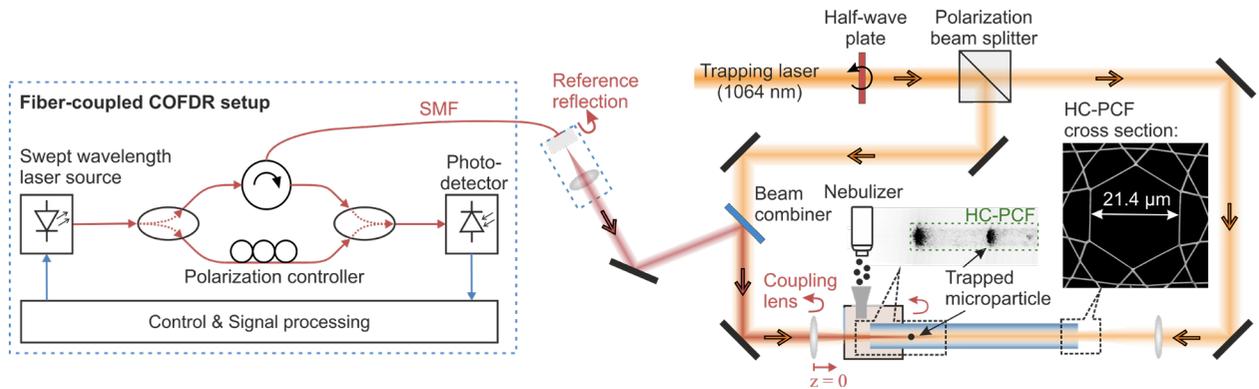

Fig. 1. Schematic of the combined COFDR (left) and particle trapping (right) setups.

It was then used to reconstruct the reflection profile as a function of the time delay between probe and reference paths. The interference visibility was maximized by adjusting the state of polarization in the reference path. The reflection profile was finally determined by calculating the fast Fourier transform (FFT). Nonlinearities introduced by the tunable laser (critical for the FFT-calculation) were compensated for by resampling the detected signal in equidistant frequency intervals using the wavelength vector recorded internally by the Luna laser source. Finally, the number of samples per peak in the reconstructed reflection profile was increased by zero padding prior to calculation of the FFT.

The frequency sweep over 5 THz results in a two-point spatial resolution of approximately 30 µm in air, given by $\Delta z = c/(2B)$ [4], where $c$ is the speed of light in a medium and $B$ the frequency tuning range. The position of a single reflector can however be resolved with significantly higher precision by least-square-fitting the measured reflection to a second order polynomial and determining the peak position from the polynomial coefficients. The positions of the reflectors are measured relative to the position of the FC/PC connector at the end of the single mode fiber (SMF), which causes a Fresnel reflection. As a result, any perturbations to the optical length of the SMF, such as those caused by thermo-optical effects, are eliminated in the measurements.

*2.2. Particle trapping setup*

The particles were captured and propelled in dual-beam trapping scheme, as illustrated in Fig. 1 (right-hand side) [5]. A stream of water droplets each containing on average a single particle, generated by a nebulizer with a mesh grid size of 4.2 µm, was directed towards the end-face of the HC-PCF inside a small chamber. Droplets were individually trapped by two counter-propagating laser beams at 1064 nm (total power of 2 W) and held in position until the water had been driven off by laser heating, leaving behind a dry particle. The trapped particle was then propelled into the hollow-core by briefly unbalancing the power ratio between the two laser beams. The experiments were conducted using polystyrene particles with diameters 1.65 µm and 4 µm, and the HC-PCF was a kagomé-style fiber with a core diameter of 21 µm (see inset in Fig. 1). An inlet tube was used to direct the droplet stream precisely towards the trapping zone. The particle loading process was monitored using a monochromatic camera (not shown in Fig. 1).

## 3. Experimental results

Position measurement to few-µm precision of particles with diameters as small as 1.65 µm was made possible by using a HC-PCF with a core diameter smaller than in previous experiments [3].

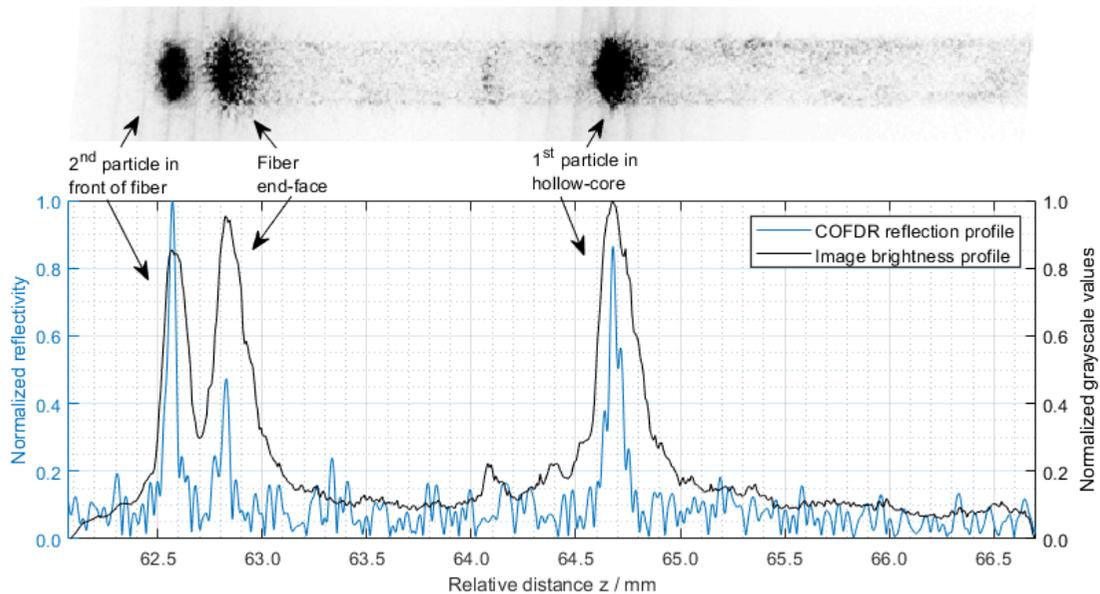

Fig. 2. Comparison of the COFDR reflection profile and the brightness profile measured by the monochromatic camera for 1.65 μm diameter polystyrene particles.

### 3.1. Positions of two particles: comparison with the camera image

In the first set of measurements, a second particle was captured in front of the fiber end-face after trapping and propelling a single particle (1.65 μm diameter) into the hollow-core. This was monitored both by the monochromatic camera (from the side) and by the COFDR system. To compare the results, the camera image was first horizontally aligned with the bright spots scattered by the particles and the fiber end-face (Fig. 2, top panel), and the brightness profile computed by integrating all grayscale values along the vertical axis. The scaling between camera pixel size and distance was estimated using the known outer diameter of the fiber as a reference. In Fig. 2 (bottom panel) the measured COFDR reflection profile (blue) is compared with the brightness profile from the camera image (black). It can be seen that both the trapped particle in the hollow-core fiber (particle 1) and the levitated particle in front of the fiber end-face (particle 2) cause highly visible peaks in the COFDR profiles. The reflection caused by the fiber end-face appears significantly lower in the CODFR reflection profile. The peak positions in both profiles are in very good agreement with each other, while the obviously narrower peaks in the COFDR measurements permit the position of the particles to be measured with much higher precision.

### 3.2. Spatial resolution

To test the resolution of the system, the COFDR measurement was repeated 75 times with two polystyrene particles (4 μm diameter) trapped at fixed positions within the hollow-core. Fig. 3a and 3b show that the positions of the two particles could be measured to standard deviations of 2.62 μm and 1.85 μm.

### 3.3. Positions of moving particles

Fig. 3(c) shows the reflection peaks measured for two polystyrene particles (4 μm diameter) at different positions along the fiber. The particles were propelled by briefly unbalancing the power ratio between the trapping beams. As expected, the parasitic reflection from the fiber end-face appears at the same position whereas the both particle reflections move by ~5.8 mm. The inter-particle distance in both COFDR measurements was 2.7 mm, varying by less than 10 μm, which indicates that the particles are optically bound as they are propelled along the fiber, as previously observed at low resolution by imaging from the side [6,7].

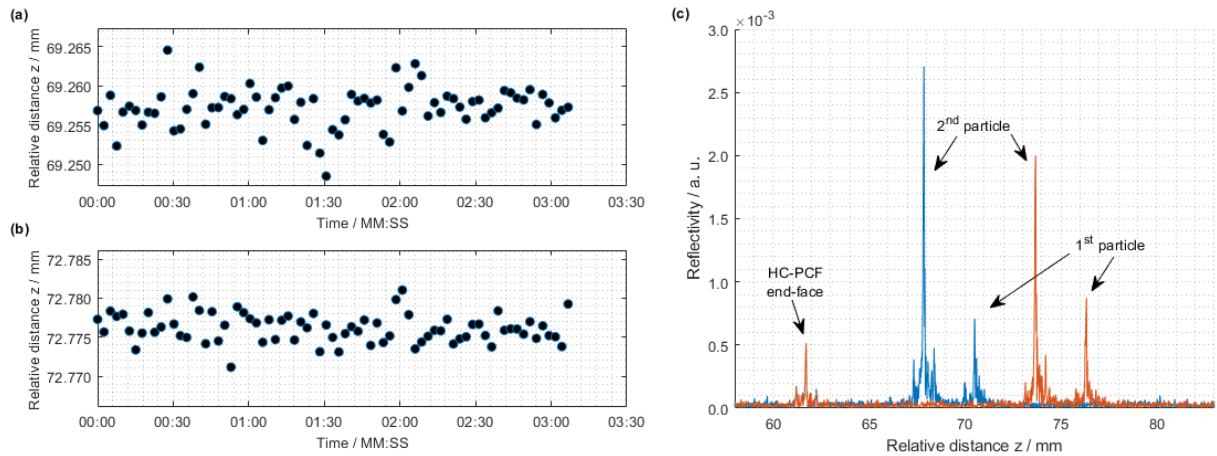

Fig. 3. (a, b) Positions (relative to the coupling lens) of the detected peaks for two optically bound polystyrene particles (4 μm diameter, spaced by 2.7 mm) over 75 repeat measurements. (c) Reflection profiles from two consecutive COFDR measurements. Two trapped particles were propelled along the HC-PCF by unbalancing the power ratio of the counter-propagating trapping beams.

## 4. Conclusions

Multiple optically trapped particles with diameters as small as 1.65 μm can be detected and accurately located in HC-PCF using COFDR. Compared to camera-based position measurement, COFDR works over much longer distances and offers much higher resolution. The results pave the way towards multi-parameter sensing with high spatial-resolution, using multiple flying particles in HC-PCF. The refresh time of the COFDR system could be further reduced by using a tunable laser with a higher sweep rate, permitting the particle dynamics to be analyzed.

## 5. Funding

This work was funded by the Deutsche Forschungsgemeinschaft (DFG, German Research Foundation) under the project number 418737652.